\documentclass[12pt,a4paper]{article}
\usepackage{amsfonts}
\usepackage{amsmath}
\usepackage{graphics}
\usepackage{graphicx}
\usepackage{amsmath}
\usepackage{multirow}
\usepackage{psfrag}
\usepackage{mathrsfs}
\usepackage{subfigure}
\usepackage{arydshln}
\usepackage{natbib}

\newcommand{\expect}{\mathsf{E}}
\newcommand{\clp}{\mathscr{P}}

\begin{document} 
\title{An optimal multi-period crossover design for an application in paediatric nephrology }
\author{J.N.S. Matthews\\
School of Mathematics \& Statistics\\
Newcastle University\\
Newcastle upon Tyne NE1 7RU, UK\\
john.matthews@ncl.ac.uk}
\date{}
\maketitle

\begin{abstract}
Crossover clinical trials can provide substantial benefits by eliminating inter-patient variation from treatment comparisons and by allowing multiple observations of each patient. They are particularly useful when sample sizes are necessarily small.  These advantages proved particularly valuable in an assessment of clot prevention in children undergoing haemodialysis. Only small numbers of children are treated at any given time in any single unit, but each patient is obliged to attend two or three times each week, suggesting the use of a crossover trial with many periods.  Standard crossover trials described in the literature a) typically have fewer than 10 periods and b) are based on a model of questionable applicability to this study. This paper describes the derivation of an optimal crossover trial with 30 periods which was used to compare the treatments using nine patients.
\end{abstract}


\section{ Introduction}

Children undergoing regular haemodialysis for renal failure need to attend the dialysis unit (DU) two or three times per week.  On each visit the dialyser is connected to the patient's circulation.  To achieve this the patient usually has a surgically inserted venous central line which remains in place permanently and is accessed on each visit to the DU.  During the interdialytic period the blood in the lumen of the line may clot.  Unclotting a line is a time-consuming problem and if it happens regularly then the central line will have to be re-sited.  Re-siting is both unpleasant for the patient and undesirable, as it is a procedure that cannot be repeated indefinitely because of the limited number of sites which can be used for this purpose.  For these patients failure to achieve reliable venous access has dire consequences.  Consequently, a small amount of anticoagulant (a ``lock'') is injected into the lumen of the central line at the end of each dialysis session in order to prevent clotting.  The usual anticoagulant is heparin (H).  This article is concerned with design issues that arose in a trial to compare heparin with alteplase (A), an alternative anticoagulant.

The dialysis treatment of choice for children with renal failure is peritoneal dialysis, but there are a few patients for whom this is unsuitable, or for whom peritoneal dialysis is no longer viable.  Consequently, in a paediatric population there are only a few children attending for haemodialysis at any centre: fewer than ten patients in a UK regional centre at any given time is typical.  Moreover, differences in treatment protocols between centres mean that a multicentre study would be difficult to run.  However, each patient who does attend the DU is obliged to do so three (or, for occasional patients, two) times a week.  Therefore information on the comparison between A and H can best be obtained using a crossover design, as this allows each patient to provide numerous observations, and at the same time improve the precision of the estimated comparison between A and H by eliminating inter-patient differences.  The trial will measure the weight of clot in the blood withdrawn from the central line at the start of each dialysis session.  As this withdrawal of blood is part of the usual clinical procedure to flush the line prior to dialysis, it imposes no burden on the patient and as such can repeated many times. 

Numerous crossover designs are available in the literature \cite[chapters 4 \& 5]{jkbook} but designs with more than ten periods are unusual in most areas of application and almost unheard of in clinical settings.  Moreover, what optimality properties these designs possess are usually based on one of a handful of rather general models which may not be the most appropriate choice for the current trial.    The next section considers the technical issues which arise in developing a design for this study, in particular the choice of a realistic statistical model for this application and its implications for design selection.  In Section \ref{sect:optdes} an optimal design is derived for this model and in Section \ref{sect:analysis} some practical issues are discussed.

\section{Choice of design and statistical model}

To avoid causing practical difficulties on the DU it was decided to run the trial for $w$ weeks, so for most patients a $3w$-period design is required but for those patients who need to attend only twice per week, a $2w$-period design is needed.  This is the first non-standard feature of the design problem: almost all ``off-the-shelf'' designs assume that each experimental subject is observed equally often.

The second issue is that most available designs assume that the response observed on subject $i$ in period $j$, $y_{ij}$, follows the model:
\begin{equation}\label{eqn:modelnocry}
y_{ij}=\mu+\xi_i+\pi_j+\tau_d(i,j)+\epsilon_{ij},
\end{equation}
where $1\le i\le n$ and $1\le j \le p$. Here $\mu$ is a general mean; $\xi_i$ is some form of subject effect, either fixed or random; $\pi_j$ is the effect of period $j$; $\tau_t$ is the effect of treatment $t$ and $d(i,j)$ is the treatment allocated to subject $i$ in period $j$, and $\epsilon_{ij}$ are independent residuals with a common variance.  More specialised designs specific to the crossover setting have been derived for a model which explicitly acknowledges the possibility of the effect of a treatment carrying over to the following period, namely
\begin{equation}\label{eqn:modelccry}
y_{ij}=\mu+\xi_i+\pi_j+\tau_d(i,j)+\gamma_{d(i,j-1)}+\epsilon_{ij}.
\end{equation}
where $\gamma_t$ is the carryover effect of $t$ and we set $\gamma_d(i,0)=0$:  see \cite[chapters 4 and 5]{jkbook} for details.

For our purposes the former model (\ref{eqn:modelnocry}) is more relevant because the effects of the anticoagulants will not carry over to subsequent periods.  This is because only a small quantity of anticoagulant is used at the end of each dialysis session.  The volume is just sufficient to fill the lumen of the central line and it largely stays {\it in situ} until the following session.  Moreover, at the start of the next session the line is aspirated, thereby removing most of the anticoagulant; any small amount left in the line will be flushed out in the course of the dialysis.  Systemic effects due to small quantities of anticoagulant diffusing from the lumen into the circulation will be negligible because i) the volume of anticoagulant involved will be very small compared with the volume of the circulation and ii) the agents will be metabolised long before the subsequent dialysis session.

Optimal designs for model (\ref{eqn:modelnocry}) are row-column designs such as Latin square and Youden designs, in which each treatment occurs equally often on each subject and in each period \cite[Chapter 4]{shahandsinha}.  These conditions are rather restrictive in that they impose conditions on the divisibility of $p$ and $n$ by the number of treatments.  Moreover, they arise from the very general form of period effect allowed by (\ref{eqn:modelnocry}),  where a separate parameter is specified for each period. 

In the present application the condition that each treatment, i.e. A and H, should be applied equally often to each patient is unlikely to be troublesome.  However, having A and H occur equally often in each period is potentially more troublesome to arrange.  The model in (\ref{eqn:modelnocry}) goes back at least to animal feeding trials \citep{cochautcan} where the appropriateness of the period effect would have been much clearer.  Its widespread adoption in the theory of crossover designs has often had a precautionary flavour - analysts feeling that designs needed to be protected again the possible untoward influences of non-specific period effects  \citep{matthewsjspi1,matthewssmmr1}, rather than that there was any clear-cut reason for the form adopted in (\ref{eqn:modelnocry}).  

In fact, the study of aspects of dialysis is an area where the inclusion of period effects in the statistical model can be justified.  As pointed out by \cite{matthoen}, many variables which can be measured during or following dialysis will be affected by the point in the dialysis cycle when the observation is made.  Most DUs operate a cycle of dialysis sessions whereby a patient attends on Mondays, Wednesdays and Fridays (or on Tuesdays, Thursdays and Saturdays but for definiteness we will assume the former, as the alternatives are equivalent).  On Mondays, variables such as the size of clots, or the amount of urea available for removal from the blood will have accumulated over the three days since attending the previous Friday, so can be expected to be different from the corresponding quantity on a Wednesday or Friday, as this will have accumulated over only two days.  

We therefore decided to adopt the following model for the expected weight of clot aspirated from a patient $i$ who attends the DU $\ell$  times per week. In period $j=1,\ldots,\ell w$.
\begin{equation}\label{eqn:realmodel}
\expect(y_{ij})=\tau d(i,j)+\pi_\ell(i,j)+\xi_i,
\end{equation}
where $d(i,j)=1$ if H is allocated to patient $i$ in period $j$ and -1 if A is allocated.  For patients attending three times per week
\begin{eqnarray}\label{eqn:periods3}
\pi_3(i,j)&=&\pi_1, \text{if } j \text{ is a Monday} \nonumber \\ 
&=&\pi_2, \text{if }j \text{ is a Wednesday}\\
&=&\pi_3, \text{if }j \text{ is a Friday.} \nonumber
\end{eqnarray}
Note that the observation $y_{ij}$ is the weight of clot aspirated following allocation of treatment $d(i,j)$ but this is only obtained when the lumen is aspirated at the \emph{following} dialysis session.  Therefore, for example, the observation $y_{i,\text{Friday}}$ which relates to the treatment allocated on the Friday, is not obtained until the following Monday.  Given that the interval between Wednesday and Friday is the same as that between Monday and Wednesday, a possible simplification  of (\ref{eqn:periods3}) is to set $\pi_1=\pi_2$. We do not adopt this slightly more prescriptive form as it transpires that this extra restriction provides no important practical advantage in the specification of the design.  

For those patients required to attend only on Mondays and Fridays we replace $\pi_3(i,j)$ by $\pi_2(i,j)$ to accommodate the different attendance schedule, namely
\begin{eqnarray}\label{eqn:periods2}
\pi_2(i,j)&=&\pi_4, \text{if } j \text{ is a Monday}  \\ 
&=&\pi_3, \text{if }j \text{ is a Friday.} \nonumber
\end{eqnarray}
The parameter $\pi_3$ is common between $\pi_2(i,j)$ and $\pi_3(i,j)$ as both groups of patients attend on Monday having attended on the preceding Friday.

The final aspect of the model which needs consideration is the residual term $\epsilon_{ij}=y_{ij}-\expect(y_{ij})$.  This represents the within-patient variation and these terms will be assumed to be independent with common variance $\sigma^2$.  The assumption of independence is questionable when each individual is observed serially over time: we will briefly consider the possible effect of dependence on the design calculations in the next section. It is also possible that the assumption of homoscedasticity will prove unfounded, although we would aim to address such issues by suitable transformation of the $y_{ij}$.

\section{Optimality calculations}\label{sect:optdes}

We suppose that $N_3$ patients dialysed three times per week and that $N_2$ patients dialysed twice weekly will be recruited, providing a total of $m=3wN_3+2wN_2$ observations.  The aim is to determine the best treatment sequences to use to define the $d(i,j)$.  A good review of this area can be found in \cite{stuf1}.  There are a variety of optimality criteria which can be used but because we are comparing only two treatments these all coincide and optimal designs are chosen to minimise the variance of the estimator of $\tau$.

\subsection{General conditions}

In order to obtain an optimal design suppose that the design matrix, $X$, implied by (\ref{eqn:realmodel}) is written as $(A \mid B_1 \mid B_2)$ where $B_1$ corresponds to the period parameters, $B_2$ corresponds to the patient parameters $\xi_i$ and $A$ is the $m\times 1$ matrix which describes the treatment allocation.  In the following we use `information matrix' to denote the inverse of the dispersion matrix, even though no likelihood is defined by the model in (\ref{eqn:modelnocry}).  If we write $\clp(X)=X(X^TX)^-X^T$ for the orthogonal projection onto the column space of $X$ and $\clp^{\perp}(X) = I-\clp(X)$, then the information matrix for $\tau$ can be written as
\begin{equation}\label{eqn:fullinf}
\mathscr{I}_1=\sigma^{-2}A^T\clp ^\perp([B_1 \mid B_2])A
\end{equation}
where $[B_1\mid B_2]$ is $X$ with the column for $A$ omitted.  Direct evaluation of this matrix is cumbersome because the number of columns in $B_2$ increases with the number of patients.  An alternative approach uses an identity for projection matrices
\begin{equation}\label{eqn:projident}
\clp^\perp([B_1 \mid B_2])=\clp^\perp(B_1)-\clp(\clp^\perp(B_1)B_2),
\end{equation}
which was first applied to crossover designs by \cite{kunert1}.  The information matrix for $\tau$ in the model which omits patient effects from (\ref{eqn:realmodel}) is
\begin{equation}\label{eqn:infred}
\mathscr{I}_2=\sigma^{-2} A^T \clp^\perp(B_1)A.
\end{equation}
Using (\ref{eqn:projident}) it can be seen that $\mathscr{I}_2-\mathscr{I}_1\ge0$ unless $A^T\clp(\clp^\perp(B_1)B_2)A=0$, in which case they coincide.  Here $A-B\ge0$ would in general mean that $A-B$ was positive semi-definite, but in the present application it is a simple inequality as $\mathscr{I}_1,\mathscr{I}_2$ are scalars.  This equation requires that $A$ should be orthogonal to the columns of $\clp^\perp(B_1)B_2$, which amounts to:
\begin{equation}\label{eqn:orth}
A^TB_2=A^T\clp(B_1)B_2.
\end{equation}  
Therefore, if we find a design which maximises the information for $\tau$ in the model with patient effects omitted \emph{and} if this design also obeys (\ref{eqn:orth}), then the design will be optimal for the full model (\ref{eqn:realmodel}).

\subsection{Detailed calculations}

In the model (\ref{eqn:realmodel}) $A$ is an $m\times 1$ vector, $B_1$ is an $m\times 4$ matrix and $B_2$ is an $m \times (N_2 +N_3)$ matrix.  In order to use the results outlined above we need expressions for $A^TB_2$, $A^T\clp(B_1)B_2$ and $A^T\clp^\perp(B_1)A$.

$A^TB_2$ is a $1 \times (N_2 + N_3)$ vector, where the $i$th element refers to patient $i$ and is the number of times patient $i$ receives H minus the number of times they receive A.  The matrix $A^T\clp(B_1)B_2$ is a $1\times (N_2+N_3)$ matrix where each element has one of two values: $q^TRP_3$ for patients treated three times per week and $q^TRP_2$ for patients treated twice a week.  Here
\[
q^T=\begin{pmatrix}q^3_M& q^3_W& q^3_F+q^2_F& q^2_M \end{pmatrix},
\]
where, e.g., $q^\ell_M$ is the number of times H was allocated on a Monday minus the number of times A was allocated on that day among patients dialysed $\ell$ times per week: corresponding differences for Wednesdays and Fridays have the subscripts $W$ and $F$ respectively. Further definitions and derivations are in the Appendix. 

The results in the Appendix, specifically equations (\ref{eqn:appr}) and (\ref{eqn:appqs}), can be combined to show that the information on $\tau$ in the reduced model (\ref{eqn:infred}) is
\begin{eqnarray*}
\sigma^{-2}A^T\clp^\perp(B_1)A&=&\sigma^{-2}\left(m -q^TRq\right)\\
&=&\sigma^{-2}\left[m-\frac{1}{w}\left(\frac{(q^3_M)^2}{N_3}+\frac{(q^3_W)^2}{N_3}+\frac{(q^3_F+q^2_F)^2}{N_3+N_2}+\frac{(q^2_M)^2}{N_2}\right)\right].
\end{eqnarray*}
A design with $q=0$, i.e. with equal replication of A and H on Mondays and Wednesdays within the thrice-weekly patients and equal replication on Mondays within the twice-weekly patients and, in addition, with $q^3_F+q^2_F=0$, will maximise $A^T\clp^\perp(B_1)A$.  Such a design would automatically mean that $A^T\clp(B_1)B_2=0$.  Therefore any such design for which $A^TB_2$ also vanishes, i.e. each patient receives A and H the same number of times, will be optimal for the model (\ref{eqn:realmodel}).  Although such designs could arise from cases where neither $q_F^2$ nor $q_F^3$ was zero, in practice it will be easier to ensure $q_F^2+q_F^3=0$ by looking for designs where $q_F^3=q^2_F=0$.

\subsection{Design Construction}\label{sect:descon}

An optimal design will provide an estimator of $\tau$ that has variance $\sigma^2/(w[3N_3+2N_2])$ and this can be used to determine the length of the trial.  The number of patients of each type (i.e. attending two- or three-times weekly) available for the study will be known at the outset.  If the usual requirements of a sample size calculation, namely an estimate of $\sigma^2$, and of the treatment difference to be detected, $2\tau_0$, then the number of weeks, $w$, needed to give a specified power and size for the associated hypothesis test can be calculated.  In the present application the precise length of the trial is not critical, so the construction of suitable designs will be expedited if $w$ is rounded up so that $w$ is even.

A clinically important difference between H and A was thought to be 10mg, giving $\tau_0=5$, and some pilot data suggested a planning estimate for $\sigma$ of 22mg.  For a two-sided 5\% test with 80\% power this gives $(\tau_0/\sigma)\sqrt{m}=1.96+0.84=2.8$, leading to $m\approx 152$.  At the planning stage there were four patients being treated three times per week and two twice-weekly, so $m=16w$ and hence a trial with $w=10$, i.e. lasting 10 weeks, is indicated.

The task is to determine appropriate sequences so that the resulting design obeys the conditions that ensure the design is optimal.  An optimal design for patients who are treated three times per week can be constructed as follows. Consider the set of four 3-sequences $S_3=\{\text{AAA},\text{AAH},\text{AHH},\text{AHA}\}$: these are all the sequences of As and Hs of length three starting with an A.  The \emph{dual} of any element of $S_3$ is the sequence with As and Hs interchanged.  Any of these sequences could be used to specify the treatments to be allocated in a particular week.  For a trial with ten weeks we need to select ten of the sequences in an appropriate way.  We proceed in two stages.  Suppose the sequences are written as in Table \ref{tab:descon1} with A-E allocated to the first five weeks and the corresponding lower-case labels to the last five weeks.  We randomly choose an element of $S_3$ and assign this to be sequence $A$ and then assign the dual of $A$ to $a$. Repeat for the remaining B-E.  In the second stage, the sequences $A$ to $e$ are randomly permuted - e.g. as in the second row of table \ref{tab:descon1}.

This construction ensures i) each patient receives each of A and H $3 \times(\tfrac{1}{2}w)$ times and that each of A and H is used equally often on each of the days Monday, Wednesday, Friday.  Clearly this method could be used for any even value of $w$.  Design sequences for patients treated twice-weekly can be constructed analogously, starting from the sequences $S_2=\{AA,AH\}$.  A design comprising arbitrary numbers of patients treated two- and three-times weekly constructed according to this algorithm will give $q=0$ and equal replication within each patient and hence will be an optimal design.

\begin{table}[!h]
\begin{center}
\begin{tabular}{|c|c|c|c|c|c|c|c|c|c|}
\hline
$A$&$B$&$C$&$D$&$E$&$a$&$b$&$c$&$d$&$e$\\
\hline
\multicolumn{10}{c}{$\downarrow$}\\
\multicolumn{10}{c}{Randomly permute}\\
\multicolumn{10}{c}{$\downarrow$}\\
\hline
$C$&$B$&$d$&$A$&$b$&$E$&$c$&$D$&$a$&$e$\\
\hline
\end{tabular}
\end{center}
\caption{Weekly sequences for the trial}
\label{tab:descon1}
\end{table}

\subsection{Dependent errors}

The error terms $\epsilon_{ij}$ are assumed to be independent, even within patients. The nature of the process under investigation in the trial, namely properties of a small volume of fluid in the lumen of an intravenous line that is flushed with many times of its own volume between each observation, makes this assumption less questionable than might often be the case with serial observations.  However, if a simple adaptation can make the design robust to dependence in the error term, then it would be prudent to use it. 

The optimal design of two-treatment crossover designs with autocorrelated error was addressed by \cite{matthewsbka}, who found that for a model with no carryover term $\gamma_{ij}$, designs with rapidly changing allocations were optimal if the autocorrelation in the residual term were positive.  For positive autocorrelation, and in the present application, this would suggest that the sequence AHA and its dual are slightly preferable to AHH and AAH, and certainly to be preferred to AAA, and its dual.  The algorithm given in Section \ref{sect:descon} did not specify the probabilities with which elements of $S_3$ are chosen.  We therefore decided to choose the elements of  $\{\text{AAA},\text{AAH},\text{AHH},\text{AHA}\}$ with respective probabilities $\{\tfrac{1}{10},\tfrac{1}{5},\tfrac{1}{5},\tfrac{1}{2}\}$.  This gives a preponderance of rapidly alternating treatments in the design, while maintaining a range of sequences. For patients treated twice per week the elements of $S_2$ can be chosen with probabilities  $\{\tfrac{1}{5},\tfrac{4}{5}\}$.  A reasonable range of sequences is desirable if the investigator wishes to leave open the possibility of performing a randomization analysis of the data, in addition to a model-based analysis.

\section{Data analysis and some practicalities}\label{sect:analysis}

Although planning for the study assumed four patients treated three times per week and two twice-weekly, by the time the trial started, seven patients treated thrice-weekly, i.e. $N_3=7$ and two treated twice weekly, $N_2=2$, were available.  Full details of the study are reported in \cite{gittins}.

There were some difficulties in the conduct of the trial.  The most important of these was that although the allocation to the treatment for the final period was done correctly, confusion on the DU meant that the corresponding observation of clot weight was missed for all thrice weekly patients.  That this observation was to have been made at the start of August, when the junior medical staff change over, may have been relevant.  In addition there are some isolated discrepancies: patient 8, one of the twice-weekly patients, missed their final four observations (three H, one A) due the availability of a kidney transplant.  Patients 1, 2 and 7 fell short of the revised total of 29 observations: patient 2 missed three observations due to holidays, while patients 1 and 7 missed one each due to errors on the DU. Although any deviation from the planned trial is undesirable, given the complexity of the design this level is probably no more than should be expected.  While the omission of a few observations will reduce the power of the study, the reasons for the missing observations should not induce any noticeable bias.  Moreover, the study was planned on the basis of $N_3=4, N_2=2$, but in fact $N_3=7, N_2=2$ were recruited, so the power would in fact have been higher than planned.

\begin{figure}
\begin{center}
\subfigure[Weight]
{
    \label{subf:rawresid}
    \includegraphics[width=7cm]{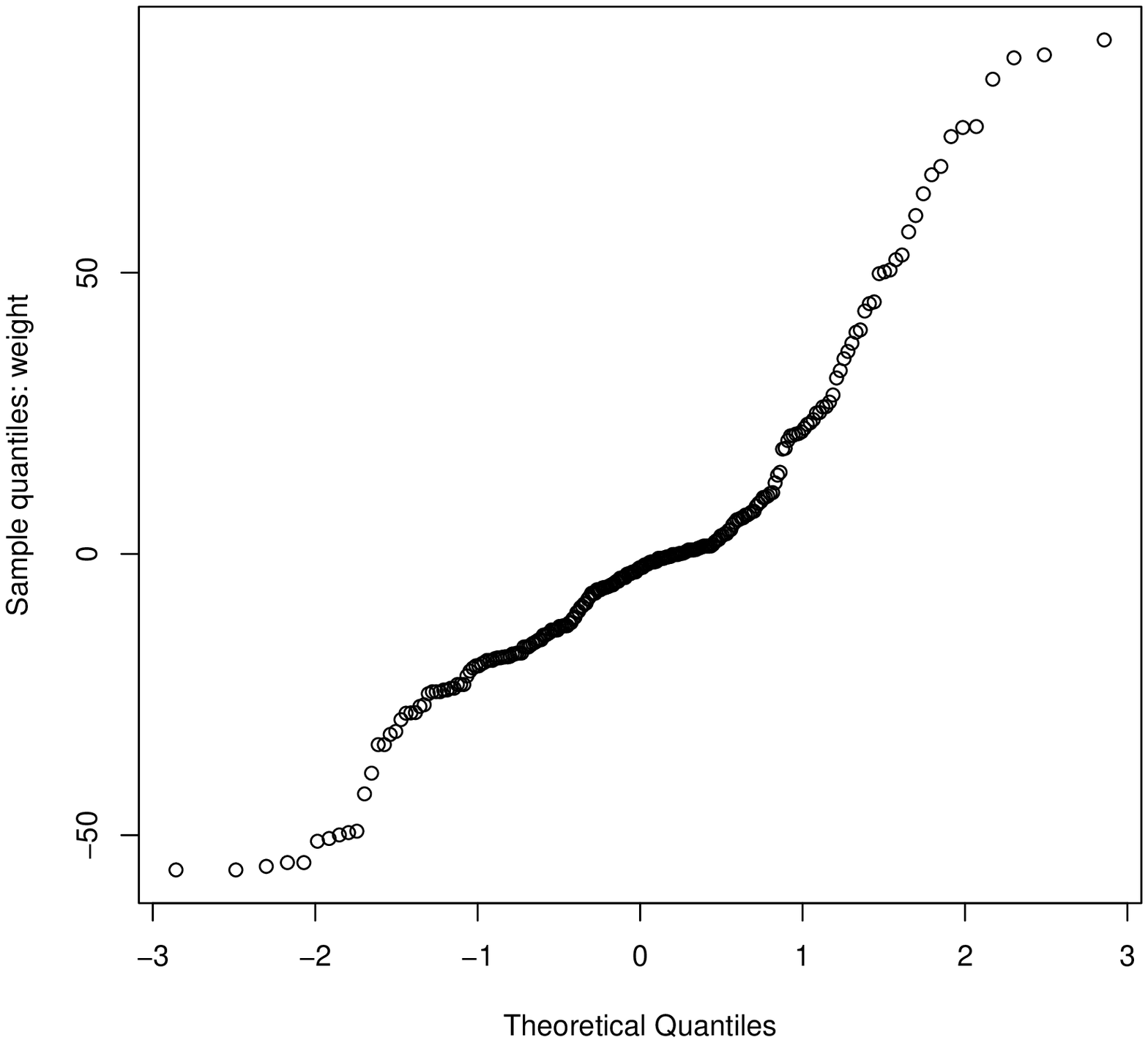}
}
\subfigure[log(Weight+0.1)] 
{
    \label{subf:loggedresid}
    \includegraphics[width=7cm]{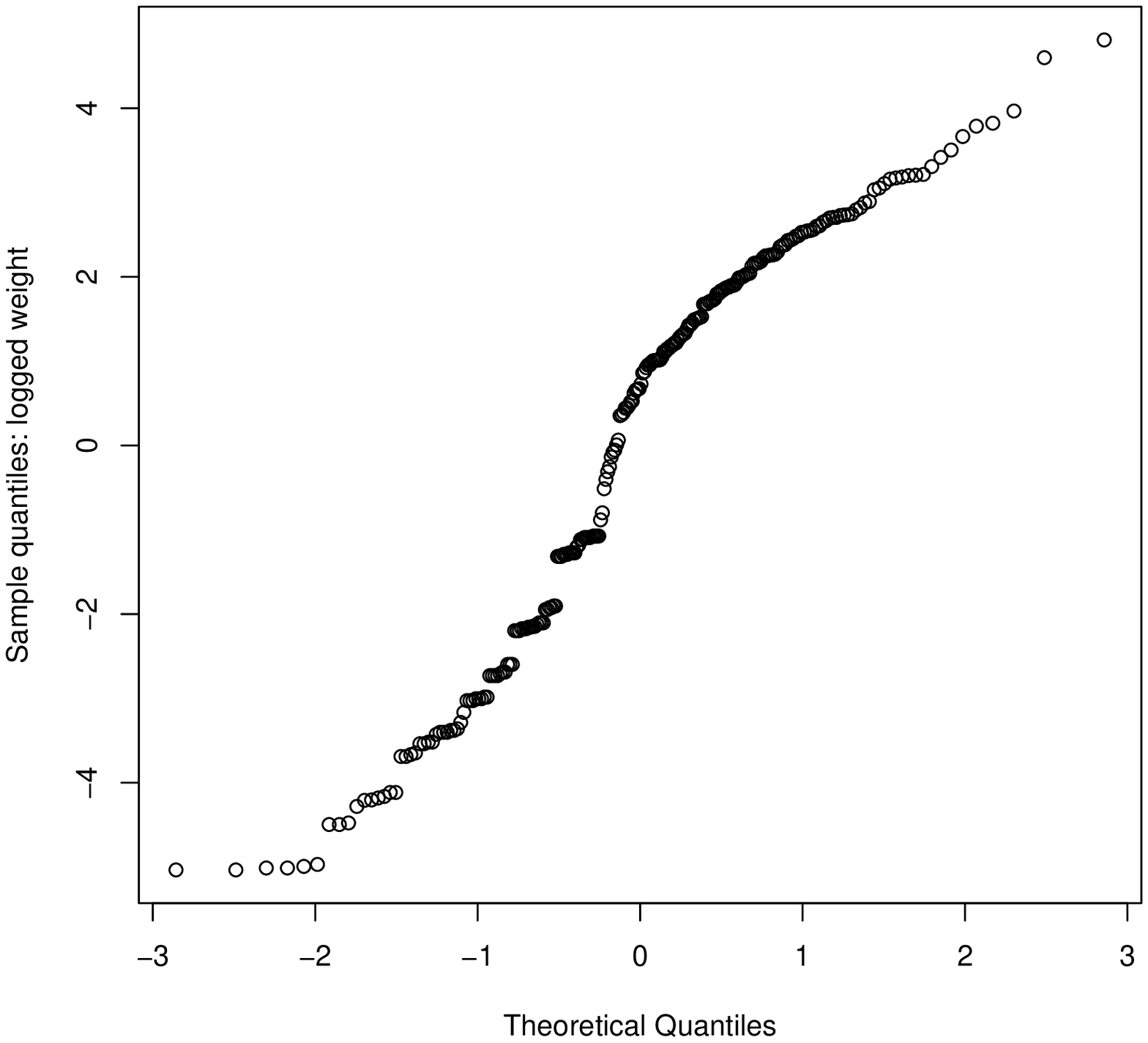}
}
\caption{Normal probability plots of the residuals from model (\ref{eqn:realmodel})}
\label{fig:qplots} 
\end{center}
\end{figure}

Fitting the model (\ref{eqn:realmodel}) gave an estimate of the mean semi-treatment difference, $\hat\tau$=2.79g, P=0.13, 95\% confidence interval (-0.79,6.37), indicating a higher mean weight of clot on heparin compared with alteplase.  A normal probability plot of residuals shown in Figure \ref{subf:rawresid} indicates that the model assumptions were questionable.  A randomization analysis based on 2000 re-randomizations of the treatment allocation using the scheme outlined in Section \ref{sect:descon} gave P=0.123, confirming the model-based P-value.   

\begin{figure}
\begin{center}
\subfigure[Heparin]
{
    \label{subf:heparin}
    \includegraphics[width=7cm]{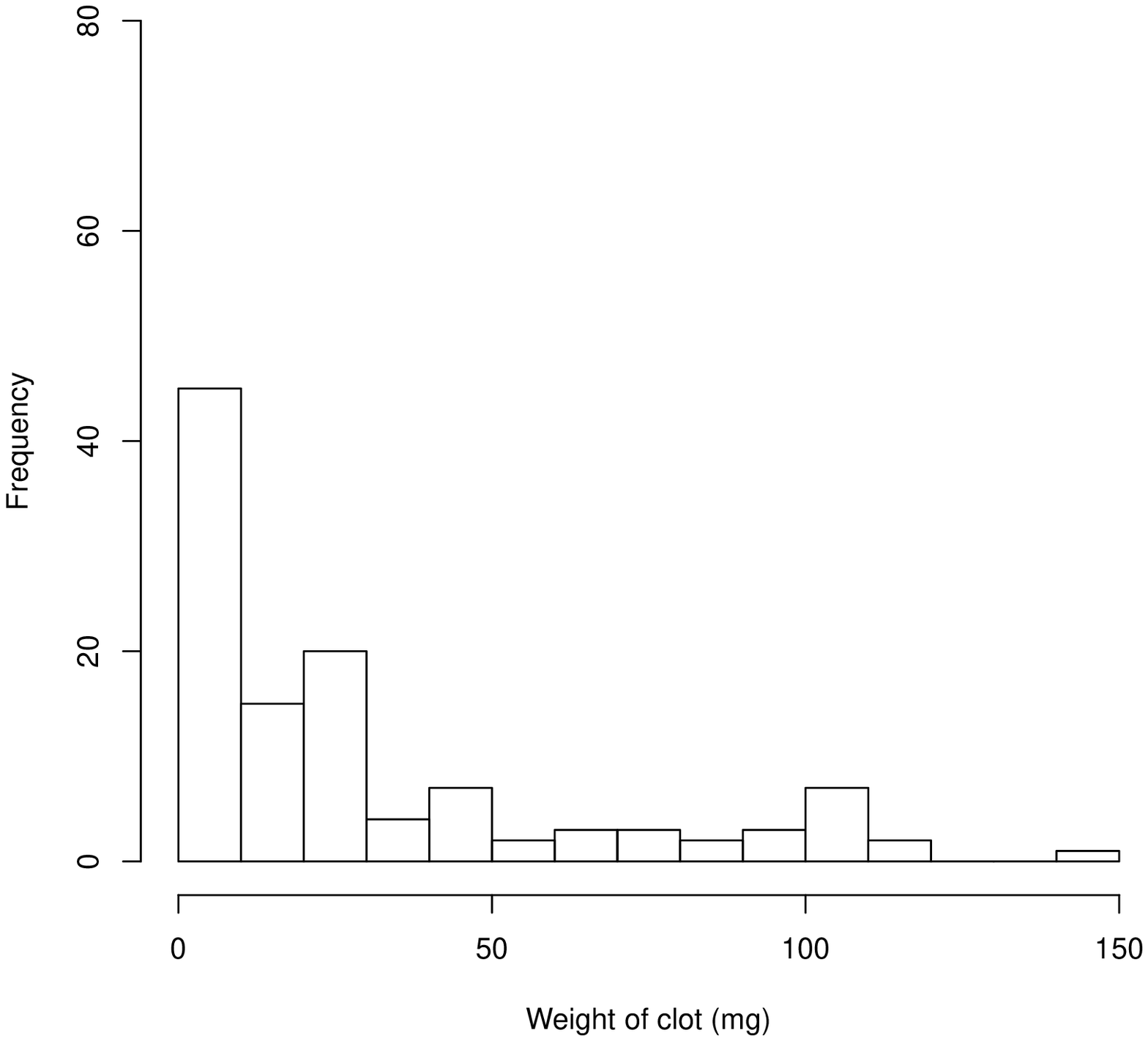}
}
\subfigure[Alteplase] 
{
    \label{subf:alteplase}
    \includegraphics[width=7cm]{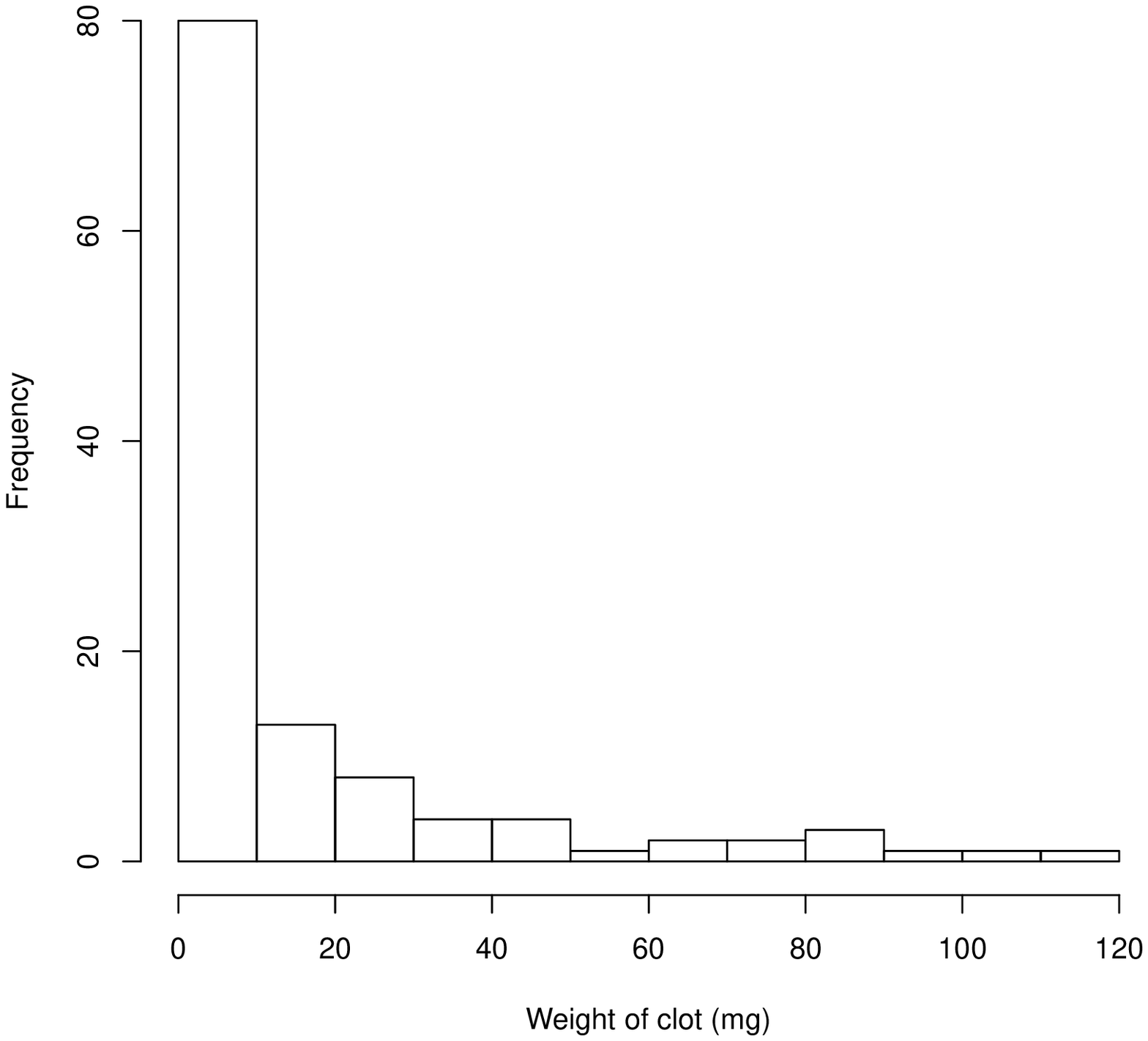}
}
\caption{Histograms of clot weight by treatment}
\label{fig:hists} 
\end{center}
\end{figure}

The probability plot in Figure \ref{subf:rawresid} indicates that the data are skewed.  This is shown, separately for each treatment, in Figure \ref{fig:hists} where it is apparent that in a high proportion of occasions (30\% H, 50\% A) no clot was retrieved from the central line.  A natural response is to skewed data is to take logs of the observations but with so many zeroes we need to use a transformation of the form $\log(y_{ij}+k)$.  Inspection of the data showed that the range of non-zero clot weights retrieved was 0.6g to 146g (H) and 0.7g to 114g (A). Rather than use a formal approach to estimating $k$, we chose $k=0.1g$.  More formal approaches using a shifted Box-Cox transformation suggested a log transformation with $k\approx0.001g$.  The estimate of the treatment effect ($2\tau$, on log scale) becomes 0.830, $P=0.013$, 95\% (0.175,1.485) ($k=0.1$) and 1.440, $P=0.018$, 95\% (0.250,2.631) ($k=0.001$).  This sensitivity to $k$, while typical of data with many zeroes, is clearly undesirable.

An alternative approach is to take the analysis in two stages.  First the probability of a clot-free sample is assessed, and then the mean weights when clot is present are analysed.  In both cases a generalized estimating equation (GEE) was used, with patient used to define the clustering variable.  The first analysis gave an odds ratio of clotting as 2.34 times larger on H than A, $P<0.001$, 95\% confidence interval (1.24,4.42). The second analysis addressed the considerable remaining skewness in the non-zero clot weights by fitting a GEE with gamma variances and a log link.  This indicated that mean clots weights were 1.86 times heavier using H than A ($P<0.001$), 95\% confidence interval (1.45,2.39).

\begin{table}[!h]
\begin{center}
\begin{tabular}{rrrrr}
&\multicolumn{2}{c}{Heparin}&\multicolumn{2}{c}{Alteplase}\\
Patient&No. of observations&\% no clot &No. of observations&\% no clot\\
\hline
1&13&46&15&67\\
2&14&29&12&58\\
3&14&7&15&40\\
4&14&43&15&93\\
5&14&43&15&20\\
6&14&21&15&33\\
7&14&7&14&36\\
\hdashline
8&7&43&9&67\\
9&10&20&10&20\\
\end{tabular}
\end{center}
\caption{Data by patient: patients 1 to 7 treated three times per week, patients 8 and 9 treated twice per week}
\label{tab:bypatient}
\end{table}

\section{Discussion}

This application required a crossover design with many more periods than is usually the case.  While multi-period designs are available from families of designs already in the literature, these require balance across patients: for example by ensuring equal replication of treatments at each period.  They also required that each patient be treated an equal number of times.  In the present application the second of these requirements would have needed the trial to be extended for the twice-weekly patients which, while possible, would have been very inconvenient for the DU.  Practical difficulties could arise with the requirement to balance over periods.  Moreover, this would arise from the use of a model for period effects that is unnecessarily general.  The use of the model (\ref{eqn:realmodel}) tailored to the application means that an optimal design can be constructed by ensuring that each patient follows a particular form of sequence, with no requirement to balance across patients.  This is a helpful feature when implementing the design.

The optimality calculations made no distributional assumptions although it did make second-order assumptions, namely that the residual terms were uncorrelated and had constant variance.  Even if these assumptions are untrue, randomization tests can provide a valid test of the hypothesis of no treatment effect.  In the event, the data proved to be rather different from that anticipated, with many instances where no clot was formed, especially when alteplase was the used as the anticoagulant.  A model-based analysis assuming Normal errors gave a $P$-value very similar to that obtained from a randomization test.  However, the conclusion of the analysis was substantially different if $\log(y_{ij}+k)$, rather that the clot weights $y_{ij}$ were analysed.   This suggests that there was also substantial heteroscedasticity in the data.  Moreover, the analyses were sensitive to the value of $k$, a parameter poorly estimated by the data.  Analyses of data of this form will seldom be adequately summarised by a single paramter such as a mean or median, so something similar to the two-stage method used will usually be necessary.

A more difficult question to address is the extent to which the properties of the design depart from optimal now that it is clear that the original assumptions are untenable.  Optimal design theory for crossover trials started with the model (\ref{eqn:modelccry}) and has developed by considering departures from this model in terms of the form of the carryover treatment term and the dependence structure of the residuals.  There is no work on design theory for crossover trials with other forms of outcome, such as a binary variable.  Designs that are optimal for the mixed binary and continuous outcome that we have pursued here would be extremely challenging to determine.

Moreover, the real issue is not how far the optimal design for the known outcomes differs from that used at the outset but could we have chosen a design that would have had good, but not necessarily optimal  properties for a range of assumptions about the outcome. There are many papers going back over many years, which derive designs which have good properties over a range of assumptions \citep{cooknacht,dumouch}, including recent contributions which address the problems posed by generalized linear models \citep{woods,steinberg}.  However, most of these contributions address robustness to mis-specification of the linear predictor or of the link function, rather than the distribution of the the outcome variable.

Although derived using optimal design theory, the present design might have been arrived at by more general considerations of orthogonality.  The design used in this study is optimal because it ensures the treatment effect is orthogonal to the period and patient effects.  Such a property can be shown to have formal advantages when the outcomes are uncorrelated with constance variance.  However, the notion that each treatment occurs equally often on each patient and on each treatment day, has more general appeal.  However, rigorous confirmation of the value of this design for the type of data that actually arose in this trial is elusive. 

\section*{Acknowledgement}

I am grateful to Drs Malcolm Coulthard and Nicola Gittins for allowing me to use their data in this paper.

\bibliography{refs}
\bibliographystyle{Chicago}

\section*{Appendix}

The matrices $B_2$ and $B_1$ are
\[
B_2=
\begin{pmatrix}
I_{N_3}\otimes 1_{3w} & 0(2wN_3,N_2)\\
0(2wN_2,N_3) & I_{N_2}\otimes 1_{2w}
\end{pmatrix}\qquad
=\begin{pmatrix}
I_{N_3}\otimes 1_{3w} & 0\\
0 & I_{N_2}\otimes 1_{2w}
\end{pmatrix}
\]
where $I_n$ is the $n \times n$ identity matrix and $1_n$ is an $n \times 1$ vector of ones, and $0(a,b)$ denotes an $a \times b$ matrix of zeroes.  Also 
\[
B_1=\begin{pmatrix}
P\\
Q
\end{pmatrix}=
\begin{pmatrix}
1_{wN_3}\otimes I_3 & 0(3wN_3,1)\\
0(2wN_2,3) & I_{wN_2}\otimes K
\end{pmatrix}
\]
where
\[
K=\begin{pmatrix}
0 & 1\\
1 & 0
\end{pmatrix}
\]
Therefore $B^T_1B_1=P^TP+Q^TQ$ can be written
\begin{equation}\label{eqn:appr}
\begin{pmatrix}
wN_3\otimes I_3 & 0\\
0 & 0
\end{pmatrix}
+
\begin{pmatrix}
0  & 0\\
0 & wN_2\otimes K^TK
\end{pmatrix}
=w\begin{pmatrix}
N_3& & & \\
 & N_3 & & \\
 & & N_3+N_2& \\
 & & & N_2
\end{pmatrix}
\end{equation}
noting that $K^TK=I_2$.

The product $B_1^TB_2$ is
\[
\begin{pmatrix}
P^T & Q^T \end{pmatrix}
\begin{pmatrix}
I_{N_3}\otimes 1_{3w}& 0\\
0 & I_{N_2}\otimes 1_{2w}
\end{pmatrix}=
\begin{pmatrix}
P^T(I_{N_3}\otimes 1_{3w}) & Q^T(I_{N_2}\otimes 1_{2w})
\end{pmatrix}
\]
Evaluating this matrix requires calculation of products such as $(1^T_{wN_3}\otimes I_3)(I_{N_3}\otimes 1_{3w})$ and this is made easier if we note that, e.g., $1_{3w}=1_w\otimes 1_3$.  We then obtain
\[
B_1^TB_2=w\begin{pmatrix}
1^T_{N_3}\otimes \begin{pmatrix}1_3\\0\end{pmatrix}&
1^T_{N_2}\otimes \begin{pmatrix}0\\0\\1_2\end{pmatrix}
\end{pmatrix}
\]

To evaluate $A^TB_1$ let $A^T=(A^T_3\;\; A^T_2)$ where $A_\ell$ is the part of $A$ relating to patients undergoing dialysis $\ell$ times per week. Then $A^TB_1=A^T_3P+A^T_2Q$ and we then obtain
\[
A^T_3P = \begin{pmatrix}q^3_M & q^3_W & q^3_F & 0\end{pmatrix} 
\] 
and
\[
A^T_2Q = \begin{pmatrix}0 & 0 & q^2_F & q^2_M\end{pmatrix} .
\]  
Here
\[
q^3_M=q^3_M(H)-q^3_M(A)
\]
where $q^3_M(H)$ is the number of times H is allocated on a Monday to a thrice-weekly patient, and $q^3_M(A)$ is the corresponding count for alteplase:  $q^3_W(\cdot)$ and $q^3_F(\cdot)$ refer to Wednesdays and Fridays, while $q^2_X(Y)$ are the corresponding quantities for patients who are dialysed twice-weekly.  It follows that
\begin{equation}\label{eqn:appqs}
A^TB_1=\begin{pmatrix}q^3_M& q^3_W& q^3_F+q^2_F& q^2_M \end{pmatrix}=q^T \text{, say}.
\end{equation}
If we write $R=w^{-1}\text{diag}\begin{pmatrix}N_3^{-1}&N_3^{-1}&(N_3+N_2)^{-1}&N_2^{-1}\end{pmatrix}$, then it follows that $A^T\clp(B_1)B_2$ can be expressed as $q^TR(B^T_1B_2)$. Each element of $A^T\clp(B_1)B_2$ corresponds to a patient and each element is either $q^TRP_3$ for thrice-weekly patients or $q^TRP_2$ for twice weekly patients, where
\[
P_3=\begin{pmatrix}1\\1\\1\\0 \end{pmatrix}\qquad P_2=\begin{pmatrix}0\\0\\1\\1 \end{pmatrix}
\]

\end{document}